\documentclass[letter]{elsarticle}

\usepackage{hyperref}
\usepackage{graphicx,caption,subcaption,amsmath,mathrsfs}
\usepackage{geometry}

\journal{Physics Letters B}
\makeatletter
\def\ps@pprintTitle{%
 \let\@oddhead\@empty
 \let\@evenhead\@empty
 \def\@oddfoot{}%
 \let\@evenfoot\@oddfoot}
\makeatother

\usepackage{xcolor}

\definecolor{darkblue}{rgb}{0,0,0.4}

\begin{document}

\begin{frontmatter}

\title{ Numerical Bootstrap in Quantum Mechanics 
}

%

\author[AffiliationA]{Jyotirmoy Bhattacharya\fnref{note}}\corref{mycorrespondingauthor}
\cortext[mycorrespondingauthor]{Corresponding author}
\ead{jyoti@phy.iitkgp.ac.in}

\author[AffiliationB]{Diptarka Das\fnref{note}}
\ead{didas@iitk.ac.in}

\author[AffiliationA]{Sayan Kumar Das\fnref{note}}
\ead{sayankumardas@iitkgp.ac.in}

\author[AffiliationA]{Ankit Kumar Jha\fnref{note}}
\ead{ankitjha539@iitkgp.ac.in}

\author[AffiliationA]{Moulindu Kundu\fnref{note}}
\ead{moulindukundu@iitkgp.ac.in}

\fntext[note]{All authors contributed equally.}

\address[AffiliationA]{Department of Physics, Indian Institute of Technology Kharagpur, Kharagpur 721302, India.}
\address[AffiliationB]{Department of Physics, Indian Institute of Technology Kanpur,
Kanpur 208016, India.}

\begin{abstract}
We study the effectiveness of the numerical bootstrap 
techniques recently developed in \citep{Han:2020bkb} for quantum mechanical systems. We find that for a double well potential the bootstrap method correctly captures non-perturbative aspects. Using this technique we then investigate quantum mechanical potentials related by supersymmetry and recover the expected spectra. Finally, we also study the singlet sector of $O(N)$ vector model quantum mechanics,  where we find that the bootstrap method yields results which in the large $N$ agree with saddle point analysis. 
\end{abstract}


\end{frontmatter}


\section{Introduction}

Recently bootstrap techniques have received a lot of attention particularly 
in the context of conformal field theories - CFTs (see the reviews \citep{Poland:2018epd} and the references therein). In the bootstrap approach one first classifies the full set of `data' necessary 
to specify a CFT completely (which are essentially the set of all operators and their correlation functions) and then constrains this data based on physical inputs such as reflection positivity, crossing symmetry and causality. It is believed that the set of constraints can be used to completely fix the operator dimensions and correlation functions in 
physical theories of high interest. For example, one of the outstanding results using 
these techniques had been to compute critical exponents of the 3D Ising model to very
high precision at much lower computational cost compared to pre-existing methods \citep{El-Showk:2012cjh}. 

Inspired by the successes of these techniques in CFTs, the authors of \cite{Lin:2020mme,Han:2020bkb} 
adapted bootstrap methods to numerically solve problems in quantum mechanics, matrix models 
and matrix quantum mechanics. Here again, the basic idea is to first identify the complete set of 
the `data' for the problem. This data a priori mainly consists of all the correlation function of the basic variables (and the eigenvalues of the Hamiltonian for the quantum mechanical 
problems). However, there are recursion relations which relate these correlation
functions. These reduce the dimension of the space of independent data significantly. 
The exact dimension of this space of independent `data' is non-universal and depends on the 
exact nature of the problem.  
Finally, this reduced space of `data' is scanned for points satisfying physical constraints 
such as positivity of the norm of all states in the system. We will denote this space of `independent data' by  $\mathscr{D}$.
In all the cases studied
so far with this technique, it has been observed that the method provides a fast convergence 
to expected results at reasonable computational costs \cite{Kazakov:2021lel} (also see \cite{Koch:2021yeb} for 
a recent use of optimization methods in multi-matrix models). 

Given the high potential of this technique in successfully 
addressing important and analytically unsolvable problems in 
matrix models and matrix quantum mechanics, it is essential to test and understand the 
effectiveness of this method for simpler and more easily tractable problems. Our work here is oriented towards this goal. In this work, we primarily follow the method outlined 
in \citep{Han:2020bkb}, and apply it to a few well chosen quantum mechanical problems. 

At first we consider the quantum double well potential problem. Here the main objective is to 
investigate to what extent this bootstrap method is able to capture the non-perturbative 
effects in the energy spectrum. We find that we are able to accurately capture the splitting of ground state energy levels due to instanton (tunneling) effects within the error bars intrinsic to the method. In the wide wall limit where the dilute gas approximation is valid, our answers match with existing analytical expressions \cite{rattazzi}.

We then consider susy quantum mechanical problems \cite{Cooper:1994eh} 
which can be reduced to a
problem of two isospectral potentials (except the ground state) that follow from the same 
superpotential. This susy system provides us with a unique opportunity to 
study and compare the convergence properties of the method, where it is known a priori that 
the spectra of two distinct potentials should converge to the same values of energy. Here we 
find that there is some difference in the convergence rates. Although the exact reason 
for this difference is not entirely clear to us, but it may possibly be attributed 
to the shifted nature of the spectrum (i.e. the ground state in one potential is the first 
excited state in the other, and so on). 

Finally, we consider the $O(N)$ vector model quantum mechanics \cite{zj}. 
The singlet sector of this model reduces to a quantum mechanical problem in collective coordinates,
which can easily be subjected to the bootstrap. This gives us the value of 
ground state energy in this system for all values of $N$ and the quartic coupling $\lambda$. 
We match the large $N$, strong coupling scaling of the ground state energy with saddle point computations carried out in the collective theory.

\underline{\bf{Note added:}} While this work was near completion, the preprint \citep{Berenstein:2021dyf} appeared which has some overlap with our work. 

\section{A brief review of the bootstrap technique introduced in 
\citep{Han:2020bkb}}\label{ssec: review}
%
Following \cite{Han:2020bkb} we shall consider a quantum mechanical system with 
the general Hamiltonian 
\begin{equation} 
H = p^2 + \mathcal V(x).
\end{equation} 
The manipulations performed in \cite{Han:2020bkb} may be summarised as follows. Choosing 
a basis of the energy eigenstates any operator $\mathcal O$ should obey the identity 
\begin{equation} \label{eqHop}
\langle [H, \mathcal O ]\rangle  = 0.
\end{equation} 
Choosing this operator to be $\mathcal O = x^s$, and using the canonical commutation $[p,x]=-i$, we obtain the identity 
\begin{equation} \label{eqqmid1}
2 \langle x^{s-1} p\rangle  - i (s-1) \langle x^{s-2}\rangle = 0 
\end{equation}  
Further choosing the operator in \eqref{eqHop} to be $\mathcal O = x^t p$ and using \eqref{eqqmid1}, we obtain another identity 
\begin{equation} \label{eqqmid2}
2 t \langle x^{t-1} p^2 \rangle + \frac{1}{2} t (t-1)(t-2) \langle x^{t-3}\rangle -  \langle x^{t} \mathcal V'(x)\rangle  = 0 
\end{equation}  
Now we can use another fact about the energy eigen states. For any operator $\mathcal O$, we must have 
\begin{equation} \label{eqHop2}
\langle H \mathcal O \rangle  =  E \langle \mathcal O \rangle ,
\end{equation} 
where $E$ is the energy eigen value corresponding to that eigen state. Using \eqref{eqHop2} for $\mathcal O = x^{t-1} $ we get 
\begin{equation} \label{eqqmid3}
\langle x^{t-1} p^2 \rangle + \langle x^{t-1} \mathcal V \rangle = E  \langle x^{t-1}\rangle 
\end{equation}  
Putting \eqref{eqqmid2} and \eqref{eqqmid3} together we get 
\begin{equation} \label{eqqmid4}
4 t E \langle x^{t-1}  \rangle  - 4 t \langle x^{t-1} \mathcal V(x) \rangle + t (t-1)(t-2) \langle x^{t-3}\rangle  - 2 \langle  x^t \mathcal V '(x)\rangle = 0 
\end{equation}  
This equation is the central recursion relation which stipulates the space of independent `data' $\mathscr{D}$ of the problem, which in turn depends on the exact nature of the potential. More precisely, $\mathscr{D}$ is determined from the first equation of the set of recurrence relation. 
Among the terms which appear in that relation, only one can be solved in terms of the others 
and the rest of the them should be considered as independent. For example, as in \cite{Han:2020bkb} if $\mathcal V = x^2 + g x^4$, this relation ensures that the `data' for the problem is $E$ and $\langle x^2 \rangle$. Whereas, if we had a harmonic oscillator the data for the problem would just be $E$. The dimension of $\mathscr{D}$ progressively increases with the 
increase in the degree of the polynomial appearing as the potential. 

The correlation functions involving momentum can be solved entirely in terms of the correlation function just involving $x$. Further all the correlation functions involving $x$ can be solved in terms of the `data' $\mathscr{D}$. Note that in all the steps above we have assumed that we are dealing with some arbitrary energy eigenstate with energy $E$.

The bootstrap is now implemented through the requirement 
\begin{equation} \label{bootop}
\langle \mathcal O^\dagger \mathcal O \rangle  \geq 0.
\end{equation} 
This requirement follows from the fact that the state must have semi positive definite norm. That is if we have a state $|\psi\rangle$ and act it with the operator $\mathcal O|\psi\rangle$, 
this new state must have semi positive definite norm. The norm of this state obtained by the action of $\mathcal O$ on $|\psi\rangle$ is simply 
$\langle \psi | \mathcal O^\dagger \mathcal O|\psi\rangle$, the above inequality follows from a constraint like this for any operator $\mathcal O$ and for any arbitrary state $|\psi\rangle$.
Again choosing the operators to be $\mathcal O = \sum_{i=0}^{K-1} c_i x^i$, \eqref{bootop} implies that the matrix 
\footnote{This matrix is refered to as the Hankel matrix as has been pointed out in \cite{Berenstein:2021dyf}.}
\begin{equation} \label{bootmat}
M_{ij} = \langle x^{i+j} \rangle 
\end{equation} 
should be positive definite. The dimension of the matrix is $\frac{K+1}{2}$, and $K$ is a variable in the procedure. The larger is the value of K the greater is the accuracy of energy eigenvalues obtained by this procedure. The entries in the matrix are $x$ correlation functions, all of which can be solved in terms of the independent `data' $\mathscr{D}$ using \eqref{eqqmid4}. Then demanding that all the eigenvalues of this matrix must be positive definite constrains 
the space of allowed data. This summarises the algorithm. 

In the following sections we shall choose different forms of the potential $\mathcal V$ and 
execute the bootstrap method to obtain some low-lying energy eigenstates. 

\subsection{The harmonic oscillator as a limit of the anharmonic oscillator}
%
%
\begin{figure*}
\centering
\begin{subfigure}{.5\textwidth}
  \centering
  \includegraphics[width=\textwidth]{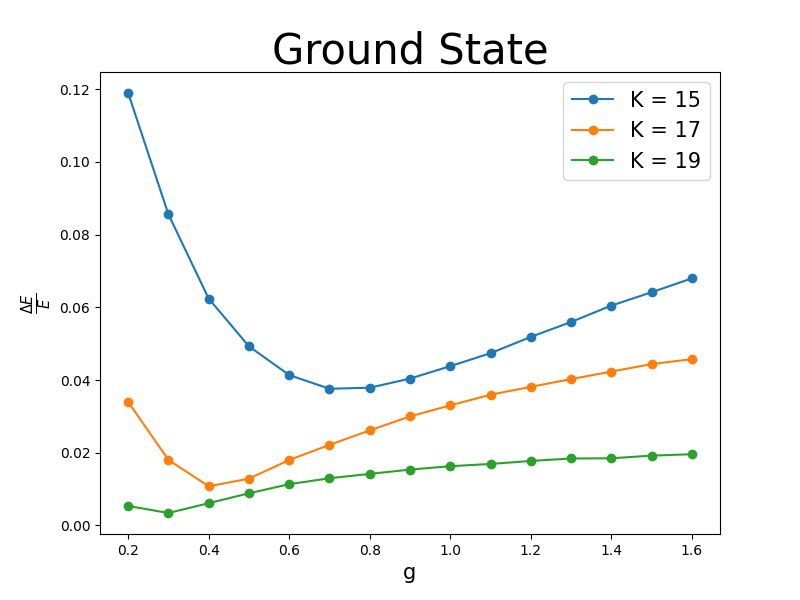}
  \caption{}  \label{fig:a1}
\end{subfigure}%
\begin{subfigure}{.5\textwidth}
  \centering
  \includegraphics[width=\textwidth]{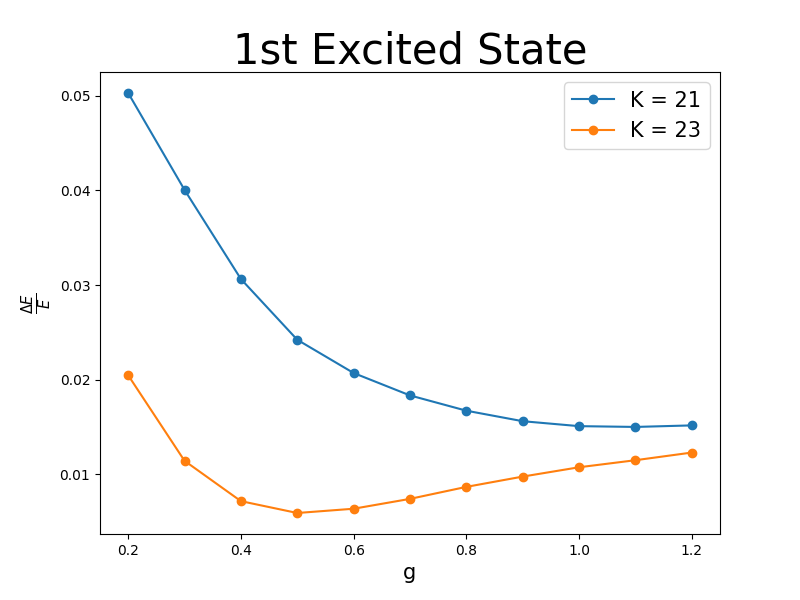}
  \caption{}  \label{fig:a2}
\end{subfigure}
\caption{Errors for the ground state (a) and first excited state (b) of the 
anharmonic oscillator, as a function of g, for a fixed value of K. We have chosen 
$m=1$ in \eqref{anplot}.}
\label{fig:a}
\end{figure*}
As a warm up of our calculation we have closely studied the anharmonic oscillator discussed in 
\citep{Han:2020bkb}. The potential is given by 
\begin{equation}\label{anplot}
\mathcal V(x) = m^2 x^2 + g x^4 . 
\end{equation}
Here we have reproduced the results and plots in \citep{Han:2020bkb}. The recursion relation 
\eqref{eqqmid4} for the potential \eqref{anplot}, implies that the data space 
is 2 dimensional for non-zero $g$, which had been taken to be $E$ and $\langle x^2 \rangle$ in 
 \citep{Han:2020bkb}. As we have noted before, if we set $g=0$ in \eqref{anplot}, the space 
 of independent data becomes one dimensional. Implementing the bootstrap for this 
 one-dimensional space is relatively easier, and one obtains the harmonic oscillator energies 
 to a very high precision \footnote{See figure 5 in \cite{Berenstein:2021dyf}. For the sake 
 of brevity we do not reproduce the figure here.}. With this background in mind, 
 we can ask what happens if we reduce the value of $g$ and study the performance of the
 bootstrap algorithm as $g \rightarrow 0$. In fig.\ref{fig:a}, we have 
 the errors as a function of $g$, for the gound state and the first excited state. We find  that for a given value of $K$, the errors generally increase as we approach $g \rightarrow 0$. Perhaps there is a  lesson in this observation for the 
bootstrap method in general. If we approach a point in the parameter space, where 
the dimensionality of $\mathscr{D}$ decreases, the accuracies are weaker. An optimal 
choice of independent `data' $\mathscr{D}$ is necessary for best performance. 

%
%
\section{The double-well potential}
%
%
\begin{figure*}
\centering
\begin{subfigure}{.5\textwidth}
  \centering
  \includegraphics[width=\textwidth]{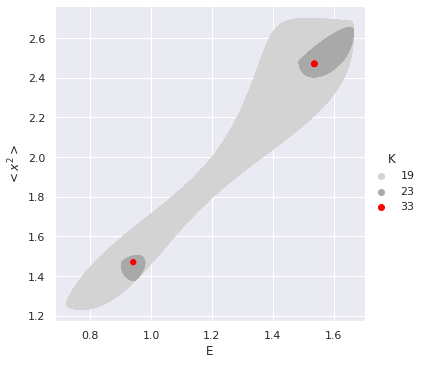}
  \caption{}  \label{fig:b1}
\end{subfigure}%
\begin{subfigure}{.5\textwidth}
  \centering
  \includegraphics[width=\textwidth]{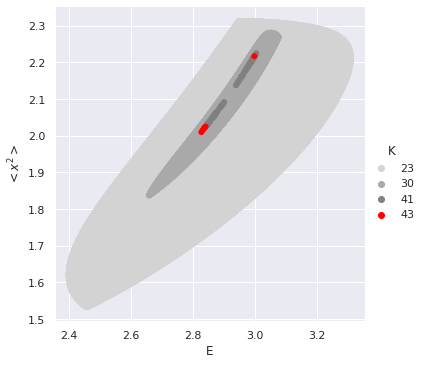}
  \caption{}  \label{fig:b2}
\end{subfigure}
\caption{In this figure we demonstrate the bootstrap method for the double-well potential
\eqref{dwpot}. In figure (a): we choose parameters to be : $m^2 = 1, ~ g = 0.2 , \mathcal V_0 = 1.25$. The two nearby lowest energy states are obtained at $E_0 = 0.942 \pm 0.001$ and $E_1 = 1.536 \pm 0.001$. Therefore, the ground state degeneracy split is $\Delta E = 0.594 \pm 0.004$. Our energy values matches with those quoted in \cite{PhysRevD.18.4767}. The degeneracy split predicted by instanton effects for these values of parameters is $\Delta E^{\text{inst}} = 1.137$. In figure (b): we choose parameters to be : $m^2 = 5, ~ g = 1 , \mathcal V_0 = \frac{25}{4}$. In this case, the two nearby lowest energy states are obtained at $E_0 = 2.834 \pm 0.007$ and $E_1 = 2.998 \pm 0.001$ and hence the ground state degeneracy split is $\Delta E = 0.16 \pm 0.01$. The instanton corrections predict $\Delta E^{\text{inst}} = 0.206$. 
}
\label{fig:b}
\end{figure*}
%
%
%
In this section we consider the double well potential and the Hamiltonian of our system is 
given by 
\begin{equation}\label{dwpot}
H = p^2 - m^2 x^2 + g x^4 + \mathcal V_0
\end{equation}
where $\mathcal V_0$ is a constant which is chosen such that 
the minima of the potential occurs at 0. 
Substituting $\mathcal V = - m^2 x^2 + g x^4$ in \eqref{eqqmid4} we obtain the following 
recurssion relation 
\begin{equation}
\begin{split}
4 t E \langle x^{t-1}  \rangle + t (t-1)(t-2) \langle x^{t-3}\rangle  +& 4 m^2 (t+1) \langle x^{t+1} \rangle \\
-& 4 g (t+2) \langle x^{t+3} \rangle
- 4 t \mathcal V_0 \langle x^{t-1} \rangle = 0 
\end{split}
\end{equation}
Due to these recursion relations in this problem, the set of independent `data' is spanned by
$\{E, \langle x^2 \rangle \}$. We now search for points in this two dimensional 
data space for which $M_{ij} = \langle x^{i+j} \rangle $ is positive semidefinite. 

Focussing on the region near the ground state we find that the condition $M \succeq 0$ is 
satisfied by two very nearby set of points, see Figs.\ref{fig:b1}, \ref{fig:b2}. We use two set 
of values of the parameters for the numerical calculation
\begin{equation}
\begin{split}
&{\text{parameter set-1 :}} ~m^2 = 1, ~ g = 0.2 , \mathcal V_0 = 1.25 \\
&{\text{parameter set-2 :}} ~m^2 = 5, ~ g = 1 , \mathcal V_0 = \frac{25}{4}.\\
\end{split}
\end{equation}
The parameter set-1 has been chosen to coincide with one of the values used in 
\cite{PhysRevD.18.4767} (see table-I in that paper). In \cite{PhysRevD.18.4767}, 
the authors perform a high precission numerical evaluation of the low-lying energy levels 
for the double well potential. Our results for first two near-degenerate levels 
matches exactly with \cite{PhysRevD.18.4767}, within the errors of our calculation,
which in some sense is intrinsic to the bootstrap method. As we increase 
the value of $K$ the errors decrease as demonstrated by the shrinking of the allowed islands in 
fig.\ref{fig:b} (see \ref{app:conv} for a study of the convergence rate of this algorithm). 

The split in the nearby ground state energy levels can be understood 
in terms of non-perturbative instanton effect. In our convention the analytical 
formula for the split in the degeneracy of the lowest energy states is given by 
\cite{Coleman:1978ae,rattazzi}: 
\begin{align}\label{split}
\Delta E^{^{\text{inst}}} &\approx \frac{ 2^{1/4} 8 m^{5/2} }{\sqrt{ \pi g }} e^{- \frac{\sqrt{2} m^3}{3 g }}.
\end{align} 
Although this is an analytical result, but it is perhaps not expected to match exactly with the  numerical results due to the approximations (dilute-gas) involved in arriving at this formula. 
Comparing the agreement of both the parameter sets (see fig.\ref{fig:b}) with \eqref{split}, 
we observe that the agreement is better when the two minimas are more widely separated.
%
\section{Susy quantum mechanics}\label{sec:susy}
%
In susy QM the problem reduces to that of two bosonic Hamiltonians once we have integrated out the fermion \cite{Cooper:1994eh}. These two Hamiltonians have the following structure 
\begin{equation} 
H_1 = p^2 + W^2(x) - W'(x) , ~ H_2 = p^2 + W^2(x) + W'(x) .
\end{equation} 
Here $W$ is the superpotential. In susy QM, the ground state energy is $E_0 =0$ for unbroken susy. 
This ground state belongs to the Hamiltonian $H_1$. There exists a correspondence between the spectrum of 
the dual Hamiltonians $H_1$ and $H_2$. This correspondence ensures that 
\begin{equation} 
E^{(2)}_n = E^{(1)}_{n+1} , \,\,\,\,\,E^{(1)}_0 = 0.
\end{equation} 
There is also a similar correspondence between the energy eigenstates
\begin{equation} 
\psi^{(1)} = \frac{1}{\sqrt{E^{(2)}_{n}}} A^\dagger \psi^{(2)}, \,\,\,\,\, \psi^{(2)} = \frac{1}{\sqrt{E^{(1)}_{n+1}}} A \psi^{(1)}
\end{equation} 
Here 
\begin{equation} 
A = \frac{d}{dx} + W(x), \,\,\,\,\, A^\dagger = - \frac{d}{dx} + W(x).
\end{equation} 
This will ensure that there is exact correspondence between the data $E$ and $\langle x^2 \rangle$ (and other higher moments $\langle x^n \rangle$, if they are a part of $\mathscr{D}$) of the two bosonic Hamiltonians, when we implement the technique of \cite{Han:2020bkb}. It is possible that the allowed regions obtained by bootstrap is distinct but has an overlap. The overlap has to exist because of the isospectrum nature of the two Hamiltonians. This overlap may be exploited to improve the convergence properties of the bootstrap 
algorithm. 

It may be observed that, even for a non-susy bosonic problem this technique may be applicable. 
The dual Hamiltonian may be constructed once we can write down the superpotential. In this case, the superpotential must be found by solving the Riccati equation 
\begin{equation} 
W^2(x) - W'(x) = \mathcal V(x) - E_0.
\end{equation} 
Note that the energy of the ground state must be subtracted from the potential to ensure the corresponding susy system has zero energy ground state. Since 
the Riccati equation is a non-linear equation it is not easy to solve it. This is the major limitation in using this method generally. 

However, choosing a specific $W(x)$ should in principle provide tractable problems involving 
potentials related by susy on which we can experiment using this new bootstrap technique. 

The most general recursion relations for the two dual Hamiltonian in terms of the superpotential
may be obtained immediately from \eqref{eqqmid4}, we get 
\begin{equation} \label{eqqmid4b}
\begin{split}
& 4 t \left( E \langle x^{t-1}  \rangle - \langle W^2(x) x^{t-1} \rangle  \mp \langle W'(x) x^{t-1}  \rangle \right) \\
& + t (t-1)(t-2) \langle x^{t-3}\rangle 
- 4 \langle W(x) W'(x) x^{t}\rangle  \mp 2  \langle W''(x) x^{t} \rangle = 0 
\end{split}
\end{equation}  
We can now choose a suitable $W(x)$ and execute the numerics. 
We shall make the following choice
\footnote{Although $\mathcal V_1$ is a double well potential, but its ground state 
has the same energy as the central maxima $E_0 = 0$. So, this potential is too shallow 
to have near degenerate states.}
\begin{equation}\label{csusypot}
W(x) = x^3 , ~ \mathcal V_1 (x) = x^6 - 3 x^2, \mathcal V_2(x) = x^6 + 3 x^2. 
\end{equation}
\begin{figure}[t]
\centering
\includegraphics[width=0.7\textwidth]{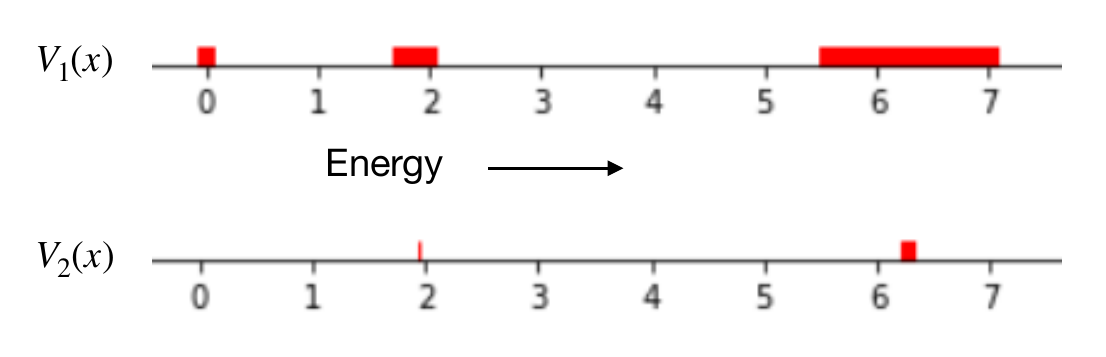}%
\caption{The allowed values of energies (marked in red) for 
potentials \eqref{csusypot} related by susy. The bootstap has been 
executed up to $K=23$.}\label{fig:c}
\end{figure}
Executing bootstrap on \eqref{csusypot} involves a 3 dimensional $\mathscr{D}$ spanned by 
$\{E, \langle x^2 \rangle, \langle x^4 \rangle\}$. 
In the region $\mathscr{D}$ which we have 
scanned, there exists two excited states apart from the ground state. In fig.\ref{fig:c} we show the spread in the allowed values in energy only, when the bootstrap is executed upto $K=23$ \footnote{For brevity and clarity we avoid showing 3D plots involving all the three parameters.}. The spread in the red lines indicate the errors in the energy levels. Within these error bars we clearly obtain a spectrum in accordance with 
the expectations from the susy arguments.  

Also from fig.\ref{fig:c} we can make a curious observation. The convergence of the 
bootstrap method is much faster for one of the potentials $V_2(x)$ compared to the 
other $V_1(x)$. In this way, 
a supersymmetric sister potential may be exploited to improve convergence of a 
given potential. Although currently we lack adequate justification of this phenomenon
\footnote{In general we have observed that the lower energy states converge 
faster than the higher energy states. For the two potentials, the ground state 
of $\mathcal V_2$ is the first excited state of $\mathcal V_1$. This shifted 
nature of the spectra may be the naive reason for the faster convergence in case 
of $\mathcal V_2$.}
, it is conceivable that a better understanding of this issue may lead to an improvement of the convergence properties of this algorithm. 

\section{The $O(N)$ vector model quantum mechanics} \label{sec:ONvec}

\begin{figure}[htbp!]
\centering
\includegraphics[width=0.52\textwidth]{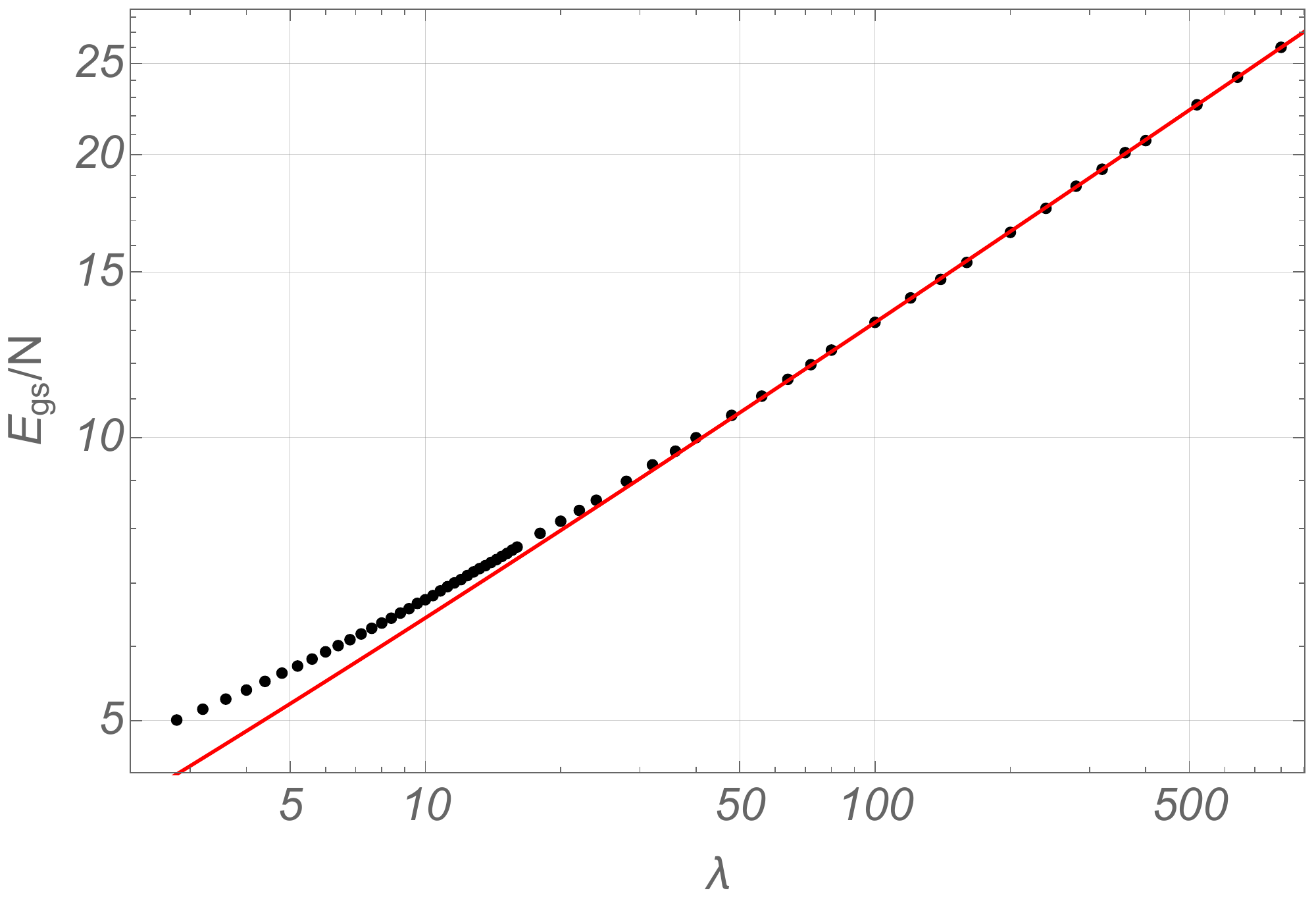}%
\caption{Ground state  energy extracted for $m=1$ and different values of the coupling $\lambda$, the straight line is the fit to $\lambda^{1/3}$. The values of energy has been obtained with a precision below 
0.4 (the precission is much higher for lower values of $\lambda$), using $K=23$. Here 
we have used $m=1$ in \eqref{effH2}.}\label{fig:on}
\end{figure}

In this section we study the $O(N)$ vector model quantum mechanics, given by the Hamiltonian 
\begin{equation}\label{vecmdl}
H = \dot{\Phi}_i \dot{\Phi}^i + m^2 \Phi_i \Phi^i + g (\Phi_i \Phi^i)^2.
\end{equation}
As is apparent from the Hamiltonian that there is a $N$ component vector $\Phi_i$, and the Hamiltonian
enjoys an $O(N)$ symmetry. If we focus on the $O(N)$ singlet sector as we do in this paper, we can move to collective coordinates \cite{Jevicki:1979mb,Das:2003vw} 
\begin{equation}
\rho = \sqrt{\Phi_i \Phi^i} 
\end{equation}
In terms of $\rho$ we can write down the following effective quantum mechanical theory for the 
singlet sector
\footnote{Note that for $g=0$, the potential \eqref{effH1} with odd values of 
$N$ are all related to eachother by supersymmetry \S \ref{sec:susy}, upto a overall additive 
constant.  } 
\begin{equation}\label{effH1}
H = \Pi^2 + \frac{(N-1)(N-3)}{4 \rho^2} + m^2 \rho^2 + g \rho^4 
\end{equation}
Here $\Pi$ is the momentum conjugate to $\rho$ and the second term arises because of the transformation of the measure due to the change
of variables. Now we can rescale $\Pi$ and $\rho$ in the following way 
\begin{equation}
\Pi \rightarrow \frac{\Pi}{\sqrt{N}} , ~ \rho \rightarrow \frac{\rho}{\sqrt{N}}.
\end{equation}
Under this rescaling the \eqref{effH1} reduces to 
\begin{equation}\label{effH2}
\frac{H}{N} =  \left( \Pi^2 + \frac{(N-1)(N-3)}{4 N^2 \rho^2} + m^2 \rho^2 +\frac{ \lambda}{4}  \rho^4 \right),
 ~ \text{where} ~\lambda = 4 g N.
\end{equation}
Once we rescale the energy as $E \rightarrow E/N$, we can bootstrap \eqref{effH2} to obtain 
the spectrum of \eqref{vecmdl}
\footnote{Note that in the effective potential \eqref{effH2}, we must have $\rho \geq 0$. But 
this condition does not affect the bootstrap procedure particularly since the 
effective potential $\mathcal V \rightarrow \infty$ as $\rho \rightarrow 0$.}. We focus on the ground state of the above Hamiltonian, and find (Fig. \ref{fig:on}) that the ground state energy exhibits a strong coupling scaling of $\sim \lambda^{1/3}$, in the large N limit.

This is consistent with analytic expectations \cite{zj}. The ground state energy can analytically be extracted at large $N$, from the low temperature limit of the partition function. In Euclidean time formulation the partition function at large $N$ is dominated by saddle point configurations, and hence $Z(\beta)$ is well approximated by $e^{-S_{E}^{\text{on-shell}}}$. Finding the saddle point configuration boils down to solving the gap equation, see appendix \S \ref{app:on} for further details. At strong coupling the ground state energy turns out to be: 
\begin{equation}
\label{scaling} E_{gs} =\frac{N}{16} \lambda^{1/3} + {\mathcal{O}}(\lambda^{-1/3}).
\end{equation}

\begin{figure*}
\centering
\begin{subfigure}{.47\textwidth}
  \centering
  \includegraphics[width=\textwidth]{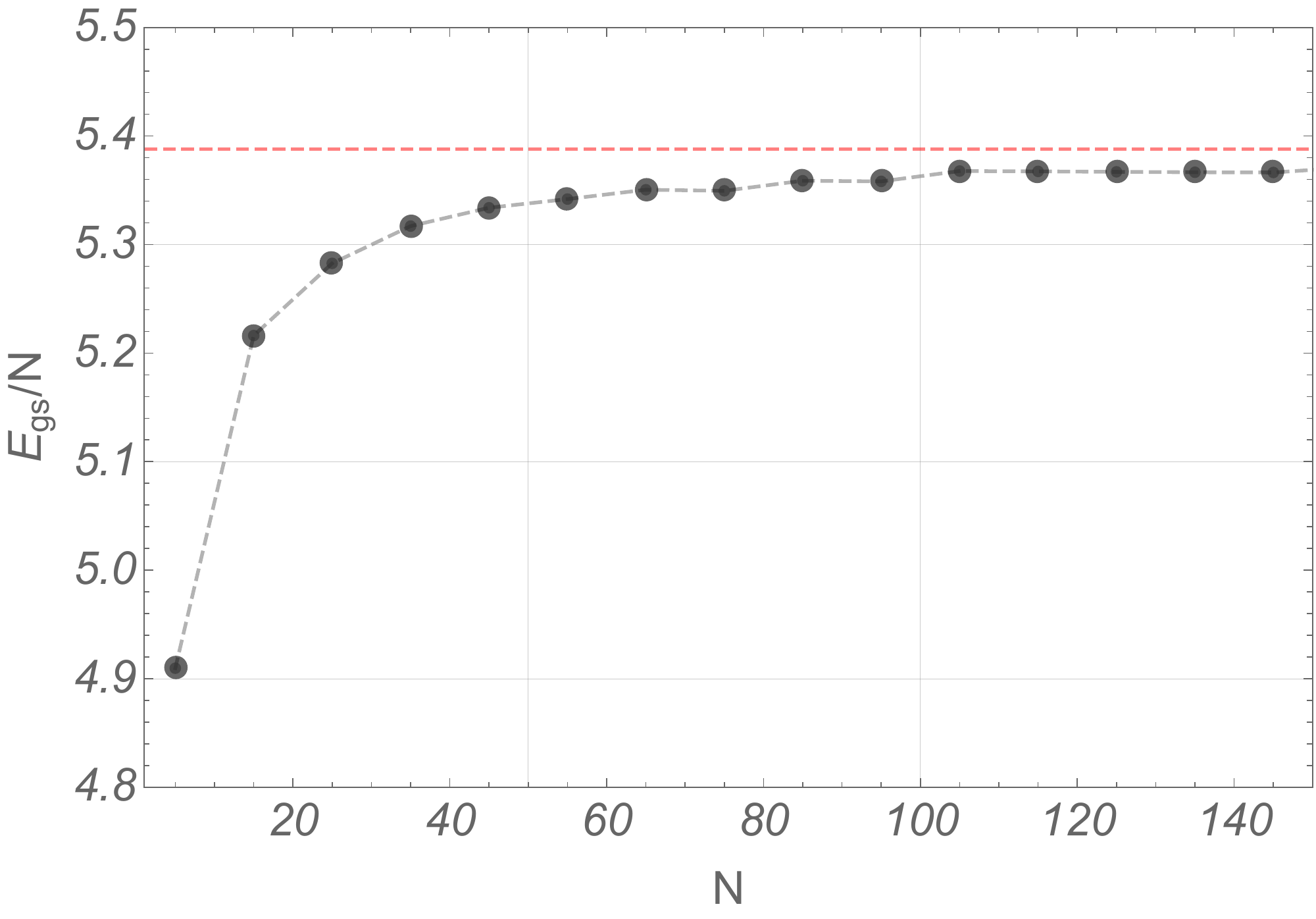}
  \caption{}  \label{fig:d1}
\end{subfigure} \quad
\begin{subfigure}{.47\textwidth}
  \centering
  \includegraphics[width=\textwidth]{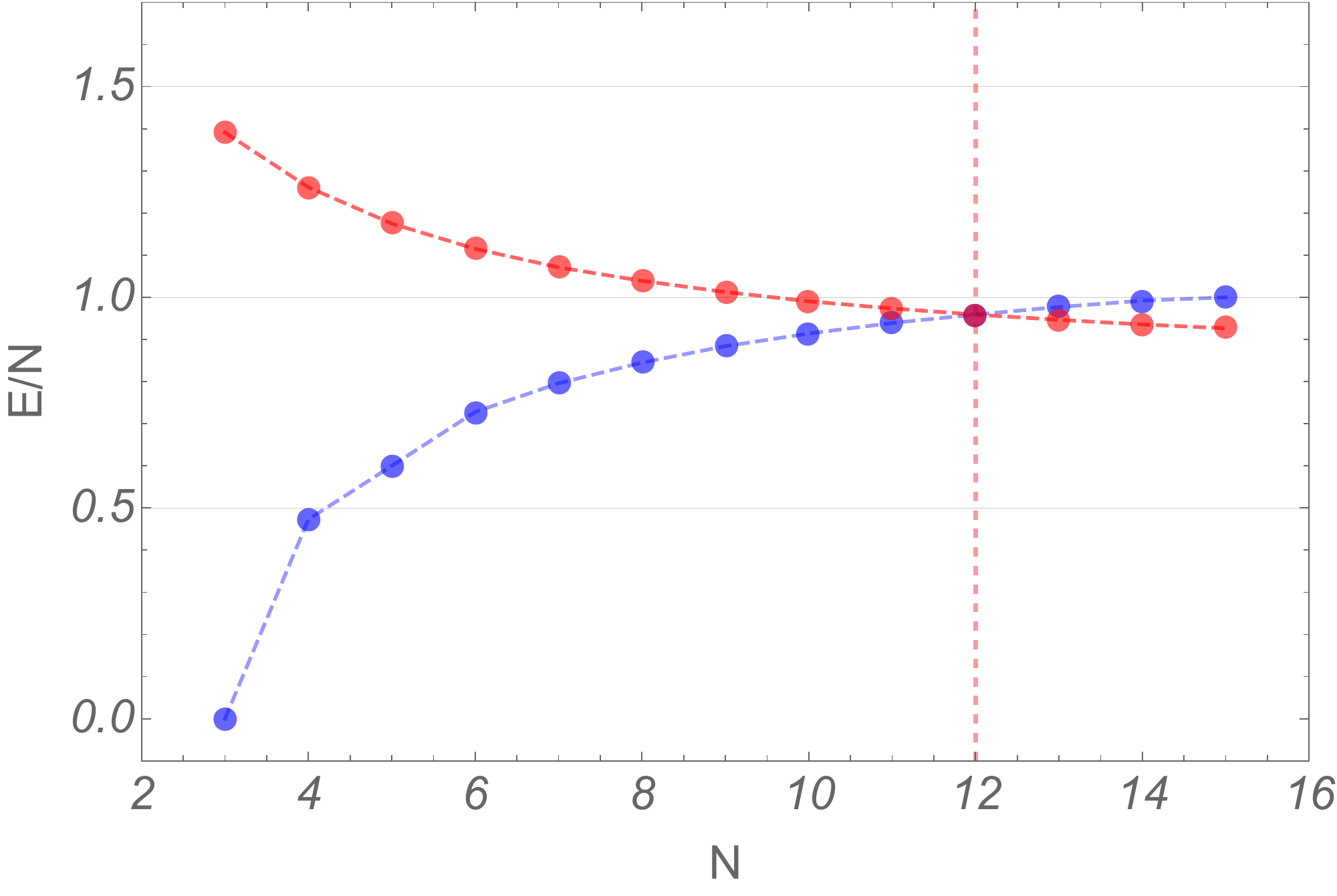}
  \caption{}  \label{fig:d2}
\end{subfigure}
\caption{Figure (a): variation of ground state energy of \eqref{effH2} with $N$. 
Figure (b): An energy level goes below the minima of the potential \eqref{effH2}
shown by the blue curve, as we increase $N$. For all values of $N \geq 12$, it remains 
as a stray allowed energy value below the minima of the potential for K values upto 25. For both of these plots we have used $m=1$ and $\lambda = 4$.}
\label{fig:d}
\end{figure*}

Note, that in this case we are working with the non-critical $O(N)$ model, where we expect the potential to remain bounded and the system to remain stable, and therefore at large $N$ identify the ground state energy with the lowest allowed island above the potential minima. In Fig.\ref{fig:d1}, we plot this normalized ground state energy as a function of $N$, and see clearly that at large $N$, $E_{gs}/N$ approaches a constant.

The gap equation for the $O(N)$ model is known to admit a critical point, when saddles coincide. However for the quantum mechanics case, the effective potential becomes unbounded from below as the critical coupling is negative, and the model only makes sense after an analytic continuation \cite{zj}. Interestingly in our bootstrap analysis, by extending the energy parameters we find an allowed island  below the potential minima and exists even at large $N$ (see \ref{app:conv} for a study of the convergence properties of this island). The crossing takes place precisely at $N=12$, see Fig. \ref{fig:d2}. Since our coupling is non-critical, we presently lack a concrete explanation for this observation, however it can well be a remnant of the instability of the critical model. 

%

\section{Discussions}

In this letter we have bootstrapped certain lamp-post examples in quantum mechanics using positivity methods of \cite{Han:2020bkb}. In the double-well example we have shown that the method captures correctly the non-perturbative instanton effects. 
The correct correspondence between the spectrum of potentials related by susy is also observed. 
The $O(N)$ vector model bootstrap also reproduces the expected low-lying spectrum
at strong coupling. 

In this work we have used a brute force method to scan the space of independent data. In 
most of the examples which we have studied, we have performed a scan with a precision of upto the third or fourth decimal place. 
This method of brute force scanning becomes progressively 
computationally costly with the increase in the dimension of the space of 
independent data. For this reason it may be 
essential to develop a more refined numerical technique to execute this search. 
The method of Lin \cite{Lin:2020mme} may be useful. It may also be advantageous to use suitably adapted gradient descent techniques, something along the lines of what was used in 
\cite{Han:2020bkb} for bootstrapping matrix quantum mechanics models 
(also see \cite{Kantor:2021jpz,Kantor:2021kbx}, where 
some new machine learning techniques have been used to yield search predictions in high dimensional spaces involving 2D CFTs). 

In our work here, it was fascinating to see how a bunch 
of inequalities lead to the exact values 
of the `data' with very high accuracies. In fact, in the case of the harmonic oscillator, we observe that for a single variable ($E$), imposing many non-linear inequalities determines it exactly to be a half-integer. This is perhaps indicative of a deeper mathematical structure 
behind the success of the method. 

Before we conclude, let us list a few other possible application of this method 
in quantum mechanical systems. 

\begin{itemize}
%
\item An important problem in 2D CFTs is the calculation of the KdV eigenvalues. This problem has been mapped to the computation of a spectral determinant of a quantum mechanical particle with potential $x^{2\alpha} + \frac{l (l+1) }{x^2}$, where $\alpha$ is a function of the central charge, $c$, and $l$ is determined in terms of the vacuum primary module as well as $c$ \cite{blz2}. The spectral determinant, $\prod_n ( 1 - E/E_n )$ is dominated by the low lying eigenvalues, and therefore the bootstrap technique may be useful to approximately determine the KdV eigenvalues. 
%
%
\item A holographic explanation for the chaotic hadronic spectrum has been the statistics of the string spectrum in a confining background, as explored, for instance in \cite{Basu:2013uva}. In the minisuperspace quantization scheme the problem reduces to coupled quantum oscillators which the authors were able to numerically solve by discretizing the Schr\"odinger eigenvalue equation. It will be interesting to find further evidence of hadronic chaos by solving for the string spectrum using the bootstrap methods in more generic backgrounds.
%
%
\item It would be extremely interesting to extend and adapt this method to finite 
temperature quantum mechanics. We have already started to explore this direction.
So far, we have been able to write down a simple 
extension of the recursion relations at finite temperature, but a tractable application 
of the numerical algorithm is proving to be significantly challenging. The primary reason 
for this difficulty is that, in this case, we must bootstrap an entire function(s) (of temperature) simultaneously rather than a few variables as we did in this paper. 
We hope to report on our progress in the future. 
\end{itemize}

\section*{Acknowledgement}
We would like to thank Pallab Basu, Shouvik Datta, Nilay Kundu, Kannabiran Seshasayanan, Vishwanath Shukla, and Dileep Jakka Pavan Surya for many useful discussions. We also thank Suchetan Das, Kannabiran Seshasayanan and Vishwanath Shukla for very helpful comments on the draft. 
We would also like to thank the anonymous referee whose suggestion led us to a more detailed  
study of the convergence rates of the bootstrap algorithm reported in \ref{app:conv}. 
DD would like to acknowledge the support provided by the Max Planck Partner Group grant MAXPLA/PHY/2018577.

%
\appendix
%

\section{$O(N)$ quantum mechanics at large $N$ } \label{app:on}

In this appendix, we find the ground state energy of the $O(N)$ quantum mechanics, by evaluating the thermal partition function, $Z(\beta)$. In the path integral formulation:
\begin{align} 
Z(\beta) &= \int [d\vec{\Phi}(\tau) ] e^{-S_E}, \,\,\,\, S_E = \int_0^\beta d\tau \left( \dot{\vec{\Phi}}^2 + m^2 \vec{\Phi}^2 + g ( \vec{\Phi}^2)^2 \right).
\end{align}
Next we rescale, $\vec{\Phi} \rightarrow \vec{\phi} = \vec{\Phi}/\sqrt{N}$ and $g N = \lambda /4$ to work with:
\begin{align}
S_E &= N \int_0^\beta d\tau \left( \dot{\vec{\phi}}^2 + m^2 \vec{\phi}^2 + \frac{\lambda}{4}( \vec{\phi}^2)^2 \right).
\end{align}
Hubbard-Stratonovich-izing the quartic term with an auxiliary field $\sigma(\tau)$ to obtain:
\begin{align}
Z(\beta) &= \int [ d\vec{\phi}(\tau) d \sigma(\tau) ] \exp \left[ - N \int_0^\beta d\tau \, \dot{\vec{\phi}}^2 + ( m^2 + \frac{\sigma}{2} ) \vec{\phi}^2 \right] \exp \left[ N \int_0^\beta d\tau\, \frac{\sigma^2}{4\lambda} \right].
\end{align}
Integrating out the vector field for the $\sigma$ dependent piece we obtain $Z(\beta) = \int [d\sigma(\tau) ] \exp[ - S_{eff} ]$ with 
\begin{align}\label{action}
S_{eff} &=  N \left[ - \int_0^\beta d\tau \frac{\sigma^2}{4\lambda} + \frac{1}{2} \int_0^\beta d\tau\, \text{Tr} \log \left( - \partial_t^2 + m^2 + \frac{\sigma}{2} \right) \right]
\end{align}
For large $N$ the above functional integral gets peaked at the saddle point, which is given by extremizing with respect to $\sigma$ (assuming it is independent of $\tau$). This gives the gap equation: 
\begin{align}\label{saddle}
- \frac{\sigma}{2\lambda} + \frac{1}{2} \int_0^\infty \frac{d\omega}{2\pi} \frac{1}{\omega^2 + m^2 + \frac{\sigma}{2} } &= 0  \,\,
\implies \frac{4\sigma}{\lambda} = \frac{1}{\sqrt{m^2 + \frac{\sigma}{2} }}.
\end{align}
For a constant $\sigma$ the finite part of \eqref{action} is:
\begin{align}
S_{eff} &= \frac{N \beta}{4}  \left( -\frac{\sigma^2}{\lambda} +\sqrt{ \frac{\sigma+ 2m^2}{2 } } \right)
\end{align}
Plugging in the solution for $\sigma$ from \eqref{saddle} into the above equation and expanding for large $\lambda$ we obtain the observed scaling \eqref{scaling}.

\section{Convergence of the algorithm}\label{app:conv}
%
\begin{figure*}[h]
\centering
\begin{subfigure}{.47\textwidth}
  \centering
  \includegraphics[width=\textwidth]{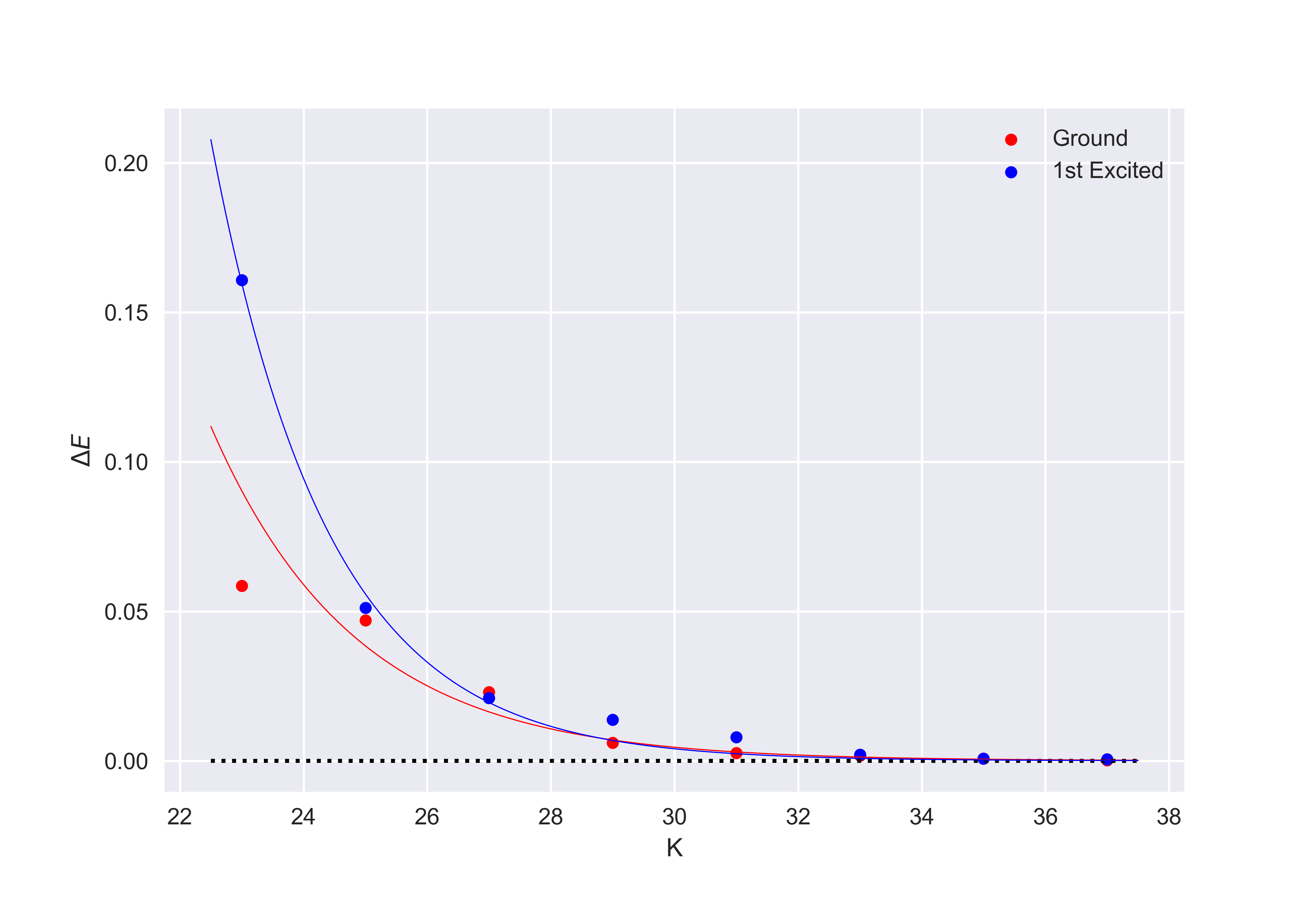}
  \caption{}  \label{fig:e1}
\end{subfigure} \quad
\begin{subfigure}{.47\textwidth}
  \centering
  \includegraphics[width=\textwidth]{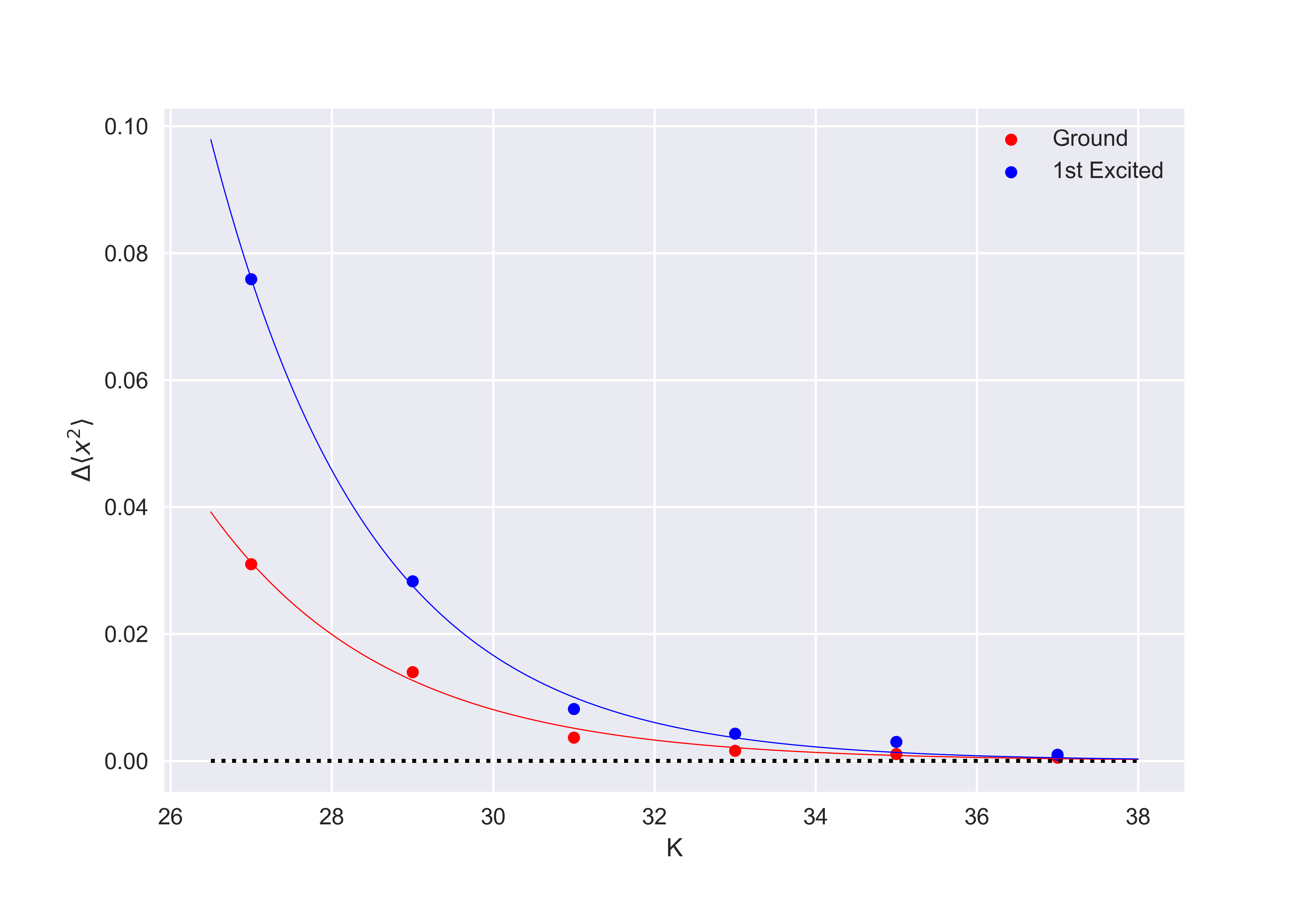}
  \caption{}  \label{fig:e2}
\end{subfigure}
\caption{ In figure (a) we plot $\Delta E$ versus $K$ for the double well potential problem with parameters identical to those in fig.\ref{fig:b1}. The solid line represents a best fit with the function 
$A \exp{\left( -\alpha K \right)}$. We have obtained $\Delta E$ for the near degenerate ground states. For the true ground state we have found $ \alpha = 0.43$, while for 
the first excited just above it $\alpha = 0.53$. In figure (b), we have plotted $\Delta  \langle x^2 \rangle$ versus $K$. Here as well we find a good agreement with the exponential fit as shown by the solid lines. The best fit values are $\alpha = 0.45, ~ 0.51$ respectively for the ground state and the first excited state. }
\label{fig:e}
\end{figure*}

We have observed that the bootstrap algorithm converges significantly rapidly in all the examples we have studied. In this appendix, we report a quantitative study of the convergence rate of this 
algorithm. 

Let us recall that the dimension of the matrix $M$ in \eqref{bootmat} is $(K+1)/2$, where $K$ is a parameter which control the accuracy of the method. 
At a given value of $K$ we obtain a set of allowed islands  in $\mathscr{D}$. All these islands shrinks in size as we increase the value of $K$. If we focus on a specific island 
the spread in energy $\Delta E = E_{\text{max}} - E_{\text{min}}$ is a good measure of the intrinsic error in this method due to the finiteness of $K$. Hence the manner in which $\Delta E$ 
decreases as we increase $K$ provides us with an estimate of the convergence rate of this method. 
A similar estimate may also be obtained from rate of shrinking of the island in other directions of $\mathscr{D}$, such as $\Delta \langle x^2 \rangle = \langle x^2 \rangle_{\text{max}} - \langle x^2 \rangle_{\text{min}}$. 

In fig.\ref{fig:e}, we have demonstrated how $\Delta E$ and $\Delta  \langle x^2 \rangle$ varies with $K$ for the double well problem with parameters identical to those in fig.\ref{fig:b1}. The analysis 
has been performed for the near degenerate ground states. The initial value of $K$ has been chosen such that the near degenerate ground states belong to distinct islands on $\mathscr{D}$. 
\begin{figure*}[hh!]
\centering
\begin{subfigure}{.47\textwidth}
  \centering
  \includegraphics[width=\textwidth]{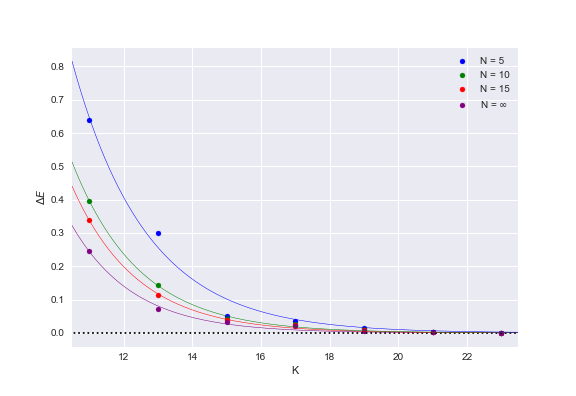}
  \caption{}  \label{fig:f1}
\end{subfigure} \quad
\begin{subfigure}{.47\textwidth}
  \centering
  \includegraphics[width=\textwidth]{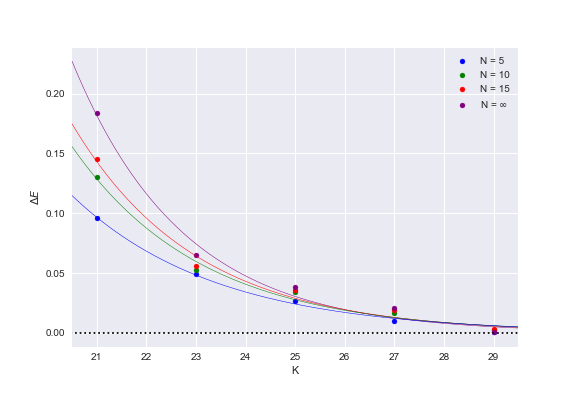}
  \caption{}  \label{fig:f2}
\end{subfigure}
\caption{In figure (a) we plot $\Delta E$ vs $K$ for the allowed island below the minima of the effective potential for the $O(N)$ model as reported in  \S \ref{sec:ONvec}. We have made 
the plot for different values of $N$ where the behaviour is very similar. The points have been fitted with $A \exp{\left( -\alpha K \right)}$ as shown by the solid lines, and the value of 
$\alpha$ in figure (a) ranges between $0.46$ to $0.55$. For comparison, in figure (b) we plot $\Delta E$ vs $K$ for the ground state above the minima of the potential corresponding 
to the same values of $N$. In this case the best fit is obtained for values of $\alpha$ ranging between $0.35$ to $0.45$.}
\label{fig:f}
\end{figure*}
We have fitted the points corresponding to $\Delta E$ and $\Delta  \langle x^2 \rangle$  for different $K$ with a function $A \exp{\left( -\alpha K \right)}$, with $A$ and $\alpha$ being the fitting parameters. 
The solid lines in fig.\ref{fig:e} represent these best fit functions, and this clearly demonstrates exponential convergence of the algorithm with the increase in value of $K$. We observe that the parameter $\alpha$ lies close to $0.5$ for both $\Delta E$ and $\Delta  \langle x^2 \rangle$. Since the dimension of the matrix $M$ is $\text{dim}(M) = (K+1)/2$, we conclude the convergence occurs 
as $\exp \left(- \text{dim}(M) \right)$. 

We then go on to analyse the convergence properties of results for the $O(N)$ vector model discussed in \S \ref{sec:ONvec}.
In fig.\ref{fig:f1}, we plot $\Delta E$ versus $K$ for the island below the minima of the effective potential for different values of $N$ (see fig. \ref{fig:d2}), and in fig.\ref{fig:f2}
we perform the same analysis for the ground state above the minima of the potential. We find that the convergence properties are very similar 
in both the cases. For this reason, we are unable to conclude whether this spurious allowed island is an artefact of the finiteness of $K$ and a physical significance 
of this point may not be immediately ruled out. We leave further investigation of this issue to future work. 
Finally, it is curious to observe that even in this case, the best fit with the function $A \exp{\left( -\alpha K \right)}$ yields values of $\alpha$ very close to $0.5$. This suggests that for the low 
energy states this bootstrap algorithm universally converges as  $\exp \left(- \text{dim}(M) \right)$.


\end{document}